\documentclass[twocolumn,aps,amssymb,showkeys,preprintnumbers]{revtex4}

\usepackage{graphicx}
\usepackage{dcolumn}
\usepackage{bm}
\usepackage{psfrag}

\begin{document}

\title{Gauge $k$-vortices}

\author{E. Babichev}
 \email{babichev@lngs.infn.it}

\affiliation{
INFN - Laboratori Nazionali del Gran Sasso, S.S. 17bis, 67010
Assergi (L'Aquila) - Italy}
\affiliation{
Institute for Nuclear Research of the Russian Academy of Sciences,
60th October Anniversary prospect 7a, 117312 Moscow, Russia}

\date{\today}

\begin{abstract}
We consider gauge vortices in symmetry breaking models
with a non-canonical kinetic term. This work extends
our previous study on global topological $k$-defects
(hep-th/0608071), including a gauge field. The model consists of a
scalar field with a non-canonical  kinetic term,
while for the gauge field the standard form of its kinetic term is preserved.
Topological defects arising in such models, $k$-vortices,
may have quite different properties as compared to ``standard'' vortices.
This happens because an additional dimensional parameter enters
the Lagrangian for the considered model --- a ``kinetic'' mass.
We briefly discuss possible consequences for cosmology, in particular, the
formation of cosmic strings during phase transitions in the early universe and
their properties.
\end{abstract}



\maketitle

\def\a{\alpha}
\def\b{\beta}
\def\e{\epsilon}
\def\o{\omega}
\def\z{\zeta}
\def\G{\Gamma}
\def\l{\lambda}

\def\pd{\partial}
\def\vp{\varphi}
\def\d{\rm{d}}

\def\be{\begin{equation}}
\def\en{\end{equation}}
\def\la{\label}

\section{Introduction}\label{SIntro}
%
Vortices are a class of topological defects which may form as a
result of symmetry-breaking phase transitions.
In condensed matter physics linear defects arise rather commonly.
Well-known examples are flux tubes in superconductors \cite{Abr} and
vortices in superfluid helium-4.
In cosmology topological defects attract much
interest because they might appear in a rather natural way during
phase transitions in the early universe. The breaking of discrete symmetries
leads to the appearance of domain walls, while the breaking
of a global or a local $U(1)$-symmetry is associated with global
\cite{glstr} and local \cite{lostr} cosmic strings, respectively.
Localized defects or monopoles may form in  gauge models
possessing a $SO(3)$ symmetry which is spontaneously broken to $U(1)$
\cite{Pol,tHo}.

Many properties of topological defects arising in symmetry-breaking models
with a canonical kinetic term are well-known,
see, e.g. \cite{Vilenkin book,Rubakov}.
Adding non-linear terms to the kinetic part of the Lagrangian
has interesting consequences for topological defects. For example,
defects can exist without a symmetry-breaking potential term \cite{Sky}.
Non-standard kinetic terms in the form of some non-linear 
function of the canonical term 
may arise in string theory, due to the
presence of higher-order corrections to the effective 
action for the scalar field. Non-canonical kinetic structures appear
also commonly in effective field theories. 

During the last years Lagrangians with
non-canonical fields were intensively studied in the cosmological
context. So-called $k$-fields were first introduced in the context
of inflation \cite{k-inflation} and then $k$-essence models were
suggested as solution to the cosmic coincidence problem \cite{k-essence,kessence others}.
Tachyon matter \cite{tachyon} and ghost condensates
\cite{ghost} are other examples of non-canonical fields in
cosmology.
An interesting application of $k$-fields is the
explanation of dark matter as a self-gravitating coherent state of
$k$-field matter \cite{Halo}. The production of
strong gravitational waves in models of inflation with
nontrivial kinetic term was considered in \cite{MukVik}.
The effects of scalar fields with
non-canonical kinetic terms in the neighborhood of a black hole were
investigated in \cite{BH}.

Recently, symmetry-breaking models with $k$-essence-like terms 
have been discussed in literature.
General properties of global topological defects appearing in such models 
were studied in \cite{kdefect}.
It was shown that the properties of such defects (dubbed $k$-defects) 
are quite different from
``standard'' global domain walls, vortices and monopoles.
In particular, depending on the concrete form of the kinetic term, the typical size of such
a defect can be either much larger or much smaller than the size of a standard
defect with the same potential term.
A detailed study of global defect solutions for one space dimension
was carried out in \cite{Bazeia}. A self-gravitating
$k$-monopole was considered in \cite{Jin}. In \cite{Adam} the
authors argued that a special type of $k$-defects may be
viewed as ``compactons'', i.e. solutions with a compact support.
Global strings with a Dirac-Born-Infeld (DBI) term were
considered in \cite{DBIstring}.

In this paper we study properties of gauge vortices arising
in a model with a $k$-essence-like kinetic term 
and a symmetry breaking potential. 
We dubb such defects ``gauge $k$-vortices'', in analogy to global
$k$-defects. 
We extend our previous investigation on $k$-defects \cite{kdefect}
including a gauge field into the model. As for the global
$k$-defects, the scalar field has a non-canonical structure
of the kinetic term, while for the gauge field we keep the canonical form
of the kinetic term. The existence
of non-trivial configurations is ensured by the symmetry-breaking
potential term. The generic feature of the model with a non-canonical 
kinetic term is the appearance of a new scale --- the kinetic ``mass''.
The presence of a new mass scale in the model radically changes 
basic properties of vortices.

We show that generally the size of the scalar core 
of the gauge vortex solution is almost independent on the presence 
of the gauge field. Its value can be approximated by the 
core' size of the global $k$-vortex with the same kinetic structure \cite{kdefect}.
With an additional, natural assumption we find that
the vector core has roughly the same size as a standard
vortex. A particularly interesting result is that 
the mass of a vortex radically vary depending on the choice of kinetic term.
As the concrete examples we study numerically the vortex solutions
for the models with DBI and with a power-law kinetic terms.

The paper is organized as follows. In Sec.~\ref{SModel}, 
we describe our model and derive its equations of motion
and the energy functional for a vortex solution. 
General properties of $k$-vortices are studied in Sec.~\ref{SGeneral}.
In Sec.~\ref{SConstraints} we find
constraints on the parameters of the model. 
Numerical solutions for particular choices of the 
non-canonical kinetic term  are presented in Sec.~\ref{Sec Numerics}.
We summarize and discuss results and cosmological applications
in the concluding Sec.~\ref{SDiscussion}.

\section{Model}\label{SModel}
%
We consider the action
\begin{equation}
  \label{act}
  S=\int {\rm d}^4 x\left[M^4 K(X/M^4) - U(f) - \frac{1}{4}F^{\mu\nu}F_{\mu\nu}\right],
\end{equation}
with
\[
  \label{X}
    X=(D_\mu\phi)(D^\mu\phi)^{*},\quad
    F_{\mu\nu}=\pd_\mu A_\nu-\pd_\nu A_\mu \,,
\]
and
\[
  \label{D_mu}
  D_\mu\equiv \pd_\mu - ieA_\mu \,.
\]
The potential term which provides the symmetry breaking is given by
\[
  \label{U}
  U(\phi)=\frac{\lambda}{4}(\mid\phi\mid^2-\eta^2)^{2} \,,
\]
where $\eta$ has dimension of a mass, while $\lambda$ is a
dimensionless constant. Note that throughout this paper we use a metric
with signature $(+,-,-,-)$. The kinetic term $K(X)$ in
(\ref{act}) is in general some non-linear function of $X$. The
action (\ref{act})  contains three mass scales: The ``usual'' 
scalar and vector masses, $\sqrt\lambda\eta$ and $e\eta$ correspondingly, 
and the ``kinetic'' mass $M$. It is worth to note
that a kinetic term that is non-linear in $X$ unavoidably leads to a new
scale in the action. In the standard case $K=X/M^4$ and the kinetic
mass $M$ drops out from the action. For non-trivial choices of the
kinetic term, the kinetic mass enters the action and changes the
properties of the resulting topological defects.

In what follows it is convenient to make the following redefinition of
variables to dimensionless units,
\[
  \label{newv}
  x\to \frac{x}{M},\, \phi\to M\phi,\, A_\mu\to M A_\mu \,.
\]
In terms of the new variables the energy density $\e$ is also dimensionless:
$\e\to M^4 \e$.
It is easy to see that $D_\mu~\to~MD_\mu$, $X~\to~M^4 X$ and the action (\ref{act}) becomes
\begin{equation}
  \label{S}
    S=\int{\rm d}^4 x
    \left[K(X)-V(\phi)-\frac{1}{4}F^{\mu\nu}F_{\mu\nu} \right],
\end{equation}
where
\begin{equation}
  \label{V}
  V(\phi)=\frac{\lambda}{4}(\mid\phi\mid^2-v^2)^{2},
\end{equation}
with $v\equiv\eta/M$ being a dimensionless quantity. One can
calculate the energy-momentum tensor from the action (\ref{S}),
\begin{eqnarray}
  \label{emt}
  T_{\mu\nu} &=& 2 K_X\mid D_\mu\phi\mid^2 - g_{\mu\nu}\left[K(X) - V(f)\right]\nonumber\\
             &-& F_{\mu\alpha}F_\nu^{\,\,\,\alpha}+
                \frac 14 g_{\mu\nu}F^{\alpha\beta}F_{\alpha\beta}\nonumber,
\end{eqnarray}
where we denoted $K_X\equiv dK/dX$, $K_{XX}\equiv d^2K/dX^2$ etc.
In the gauge $A_0=0$, the energy density for a static configuration,
$\dot\phi=0$, $\pd A_i=0$, is
\begin{equation}
  \label{E}
  T_0^0=- K(X) + V(\phi)+\frac 14 F_{ij}^2.
\end{equation}
Note also that for static configurations $X=-D_i\phi (D_i\phi)^*$.
The mass per unit length of a vortex, $E$, can be expressed as:
\begin{equation}
  \label{Ef}
    E=\int\left[-K\left(-|D_i\phi|^2\right)+V(\phi)+\frac{1}{4}F^{\mu\nu}F_{\mu\nu}\right]
	{\rm d}^2x.
\end{equation}
From the variation of the action (\ref{S}) with respect to $\phi^*$ and
$A^\mu$ we obtain as equations of motion (EoM)
\begin{eqnarray}
    \label{eom phi}
    K_X D_\mu D^\mu\phi + K_{XX} X_{,\mu}D^\mu\phi + \frac{dV}{d\phi^*} &=& 0,\\
    \label{eom A}
    \pd_\mu F^{\mu\nu} &=& e j^\nu \,,
\end{eqnarray}
where the current $j_\mu$ is given by
\[
    \label{current}
    j_\mu = -iK_X\left[\phi^*D_\mu\phi-\phi\left(D_\mu\phi\right)^*\right].
\]
(Note the additional $K_X$ in the above expression as compared to the standard case.)
One can check that the current $j_\mu$ is conserved,
\[
    \pd_\mu j^\mu=0,
\]
similar to the standard case.
To obtain the solution describing a vortex we use the
following ansatz,
\begin{eqnarray}
  \label{ansatz}
    \phi(x) &=& e^{i\theta} f(r),\\
    A_i(x)  &=& -\frac{1}{e r^2}\epsilon_{ij} r_j\alpha(r)\nonumber \,,
\end{eqnarray}
where $r=\left(x^2+y^2\right)^{1/2}$.
It is worth to note that we use the same ansatz (\ref{ansatz}) as in the standard case.
Substituting (\ref{ansatz}) into (\ref{eom phi}) and (\ref{eom A}) we obtain
the EoM for the functions $f(r)$ and $\alpha(r)$,
\begin{eqnarray}
    -K_X\left[\frac1r \frac{d\,(rf')}{dr}-\frac{f(1-\alpha)^2}{r^2}\right]\quad\quad\quad\quad
        &&  \nonumber\\
        -K_{XX}X'f'+\frac{\lambda}{2}\left(f^2-v^2\right)f&=&0,\label{EOM}\\
    -\frac{d}{d r}\left(\frac{\alpha'}{r}\right)
        - \frac{2 e^2 f^2}{r}(1-\alpha)K_X &=& 0. \label{EOMa}
\end{eqnarray}
One can check that in the standard case, $K(X)=X$,
Eqs.~(\ref{EOM}) and (\ref{EOMa}) take the familiar form
of EoMs for ``usual'' vortices.

We assume that the kinetic term $K(X)$ has the standard
asymptotic behavior at small $X$. This means that in the
perturbative regime in trivial backgrounds
the dynamics of the considered system
is the same as with a canonical kinetic term. This
requirement is introduced to avoid troubles at $X=0$: In the case of
$X^\delta$ with $\delta<1$ there is a singularity at $X=0$, while  for
$\delta>1$ the system becomes non-dynamical at $X=0$ \cite{kdefect}.
One can understand this better in terms of an ``emergent geometry'':
Because of the non-linearity of the EoMs, small fluctuations of the
scalar field feel an ``effective'' metric, which in general differs from the
gravitational one (in our case --- from the Minkowski metric).
As it was shown in \cite{causality}, in the case when a kinetic term does not
coincide with the canonical one (probably, up to some constant) in
the limit of small $X$, the effective metric for small perturbations
diverges as $X\to 0$. This means that such models are physically meaningless.

In the opposite limit, $X \gg 1$, we restrict our attention to the
modifications of the kinetic term having the following asymptotic,
\[
  \label{pow}
  K(X)=-\left(-X\right)^n.
\]
Note that for a static configuration $X<0$ and a minus sign
in the expression above  for $K(X)$ provides a positive contribution 
to the energy density (\ref{E}).
Below we find the criteria for the Lagrangians to have the desired
asymptotic $X\gg 1$ in the core of a vortex. Summarizing,
we will consider kinetic terms with the following asymptotic behavior,
\begin{eqnarray}
  \label{model}
  K(X)=\left\{
  \begin{array}{lcl}
    X,&& X\ll 1,\\
    -\left(-X\right)^n, && X\gg 1.
  \end{array}
  \right.
\end{eqnarray}
Assuming $X\gg 1$, one can easily obtain
from (\ref{model}), (\ref{EOM}) and (\ref{EOMa})
EoMs in this regime,
\begin{eqnarray}
    \left[\frac1r \frac{d\,(rf')}{dr}-\frac{f(1-\alpha)^2}{r^2}\right]
        +(n-1)\left(\ln X\right)'f'&&  \nonumber\\
        -\left(-X\right)^{1-n}\frac{\lambda}{2n}\left(f^2-v^2\right)f \!\!&=&\!\! 0,\label{eomnf}\\
    \frac{d}{d r}\left(\frac{\alpha'}{r}\right)
        + \frac{2 n e^2 f^2}{r}(1-\alpha)\left(-X\right)^{n-1}\!\! &=&\!\! 0.\label{eomna}
\end{eqnarray}
As a particular example we choose a DBI-like kinetic term for the scalar field,
\begin{equation}
  \label{BI}
  K(X)=1-\sqrt{1-2X}.
\end{equation}
It is easy to see that for this choice in the limit $X\gg 1$
the kinetic term is of form (\ref{model}) with $n=1/2$. Another
particular example we will study is a power-law form of 
the Lagrangian:
\begin{equation}
  \label{X3}
  K(X)=X+X^3.
\end{equation}
%
\section{General properties}\label{SGeneral}
%
The EoMs (\ref{eomnf}), (\ref{eomna}) for arbitrary $K(X)$
are highly non-linear and cannot
be solved analytically. We restrict our attention to the study of
vortices arising from Lagrangians having the asymptotic
behavior (\ref{model}) for the kinetic term. Although the EoMs can not
be integrated even in this case, some 
general features can be extracted without the knowledge of explicit
solutions.

At some point of our estimations we will use an additional 
simplifying assumption, which makes the understanding of the results 
more transparent. 
We will assume that the parameters of the Lagrangian
satisfy the following natural relation,
\begin{equation}
    \label{1mass}
    e\sim\sqrt\lambda,
\end{equation}
which means that in the linear regime for the kinetic term, $K(X)=X$,
the ``scalar'' and ``vector'' masses are of the same order.
Thus the assumption (\ref{1mass}) reduces the number of 
different scales in the model from $3$ to $2$:
one is the usual ``scalar'' or ``vector'' mass, $e\eta\sim \sqrt\lambda\eta$, 
and the other is a new kinetic mass $M$.
%
\subsection{The region $r\to 0$}
%
We start our study from the region close to the center of a vortex, $r\to 0$.
As we assume that in the core of a defect the kinetic term can
be approximated by (\ref{model}), we must require that for models
(\ref{BI}) and (\ref{X3}), $X\gg 1$. In the opposite case, $X\lesssim 1$,
we end up with a solution which does not deviate much from the standard
one. For $r\to 0$  we search a solution in the following form,
\begin{eqnarray}
    \label{r0}
    f(r) &=& A_f r+B_f r^3+ O(r^5),\nonumber\\
    \alpha(r) &=& A_\alpha r^2 + B_\alpha r^4+ O(r^6) \nonumber
\end{eqnarray}
with unknown constants $A_f$, $B_f$, $A_\alpha$, $B_\alpha$.
Substituting the above expressions into
(\ref{eomnf}) and (\ref{eomna}) we find that
$A_f$ and $A_\alpha$ are arbitrary, while the others
are
\begin{eqnarray}
  \label{appr}
    B_\alpha &=& -2^{n-3}e^2 n A_f^{2n},\nonumber\\
    B_f      &=& -\frac{2^{n-5}}{n^2}\lambda v^2 A_f^{2n-1}
        +\frac{n(n-2)}{4}A_f A_\alpha.\nonumber
\end{eqnarray}
The standard asymptotics for $K(X)=X$ are recovered  from the 
above expression by setting $n=1$. Note that the constants $A_f$ and
$A_\alpha$ are left undetermined, which means that the size of a
defect and its mass are undetermined too. It is possible, however,
to estimate these quantities without solving explicitly EoMs, as we will
see in \ref{subs structure} and \ref{subs mass}.
%
\subsection{Structure of a vortex}\label{subs structure}
%
The model (\ref{act}) contains a complex scalar field and a gauge
vector field. In accordance to this, there are two distinct cores for a  
vortex solutions: One is associated with the scalar
field and the other with the vector field.
The typical sizes of the cores depend on the parameters of the Lagrangian.
In this subsection we will estimate the typical sizes of cores without
explicitly solving EoMs.
In what follows it will be helpful to use an additional rescaling:
\begin{equation}
  \label{rescaling}
    f\to vf,\quad r\to r\,L_H,
\end{equation}
where
\begin{equation}
  \label{LH}
    L_H=v\;\e^{-1/2n},
\end{equation}
and $\e$ was defined as 
\begin{equation}
  \label{eps}
    \e\equiv \lambda v^4,
\end{equation}
in analogy with global defects \cite{kdefect}. Later we will
see that the quantity $\e$ (\ref{eps}) corresponds to the energy density inside
the scalar core of a vortex. The rescaling (\ref{rescaling}) brings
the EoMs (\ref{eomnf}), (\ref{eomna})  to the following form,
\begin{eqnarray}
    \left[\frac1r \frac{d\,(rf')}{dr}-\frac{f(1-\alpha)^2}{r^2}\right]
        +(n-1)\left(\ln X\right)'f'&&\quad\quad \nonumber \\
        -\frac{1}{2n}\left(-X\right)^{1-n}\left(f^2-1\right)f &=& 0, \label{eomHf} \\
    \frac{d}{d r}\left(\frac{\alpha'}{r}\right)
        + \frac{2n}{\gamma}\left(-X\right)^{n-1}\frac{f^2(1-\alpha)}{r} &=& 0, \label{eomHa}
\end{eqnarray}
where
\begin{equation}
    \label{gamma}
    \gamma  
        = \frac{\lambda}{e^2}\;\e^{-2(n-1)/n}.
\end{equation}
Notice that in the standard case, $n=1$, the parameter $\gamma$ 
coincides with the usual parameter, defined as the ratio of the scalar and vector masses.
The EoM for $f(r)$, Eq.~(\ref{eomHf}), contains parameters of order of $1$ as 
well as the function $\alpha$, going from $0$ to $1$.
Therefore one can guess that the typical scale on which the function
$f(r)$ varies is of the order
$L_H$. Thus the typical size of the scalar core, $l_H$, is given by
\begin{equation}
    \label{lH}
    l_H \sim L_H,
\end{equation}
and is almost independent on $\gamma$ and $e$. In fact,
the value $l_H$ (\ref{lH}) coincides
(up to an irrelevant numerical factor of order of $1$) with the
size of the core in the case of a global $k$-string, found in \cite{kdefect}.
Very roughly speaking, the scalar core remains unaffected by the gauge field.
Intuitively it can be understood as follows. Topological defects exist due to
the presence of a potential, which provides symmetry breaking for the
scalar field $\phi$; the gauge field is in a sense merely an auxiliary
component.
Therefore, the presence of the gauge field should not
radically change the size of the scalar core.

To estimate the size of the vector core is a more tricky task.
First of all we note that when $\gamma\sim 1$, i.e.
\begin{equation}
    \label{valideq}
    \frac{e}{\sqrt\lambda}\;\e^{(n-1)/n}
                                                \sim 1,
\end{equation}
the size of the vector core, $l_V$, is of order of the size of the scalar core,
\begin{equation}
    \label{lVeq}
    l_V\sim l_H,
\end{equation}
since in this case the EoMs (\ref{eomHf}), (\ref{eomHa}) do not
contain any large or small parameters. 

To consider other cases, when the vector core is much larger/smaller
than the scalar core, let us turn back to Eq.~(\ref{eomna}).
The inflection point
for the function $\alpha(r)$ is at the point $r\sim l_V$. Then, taking also into account
that $\alpha'(r)\sim 1/l_V$ at $r\sim l_V$, we obtain from (\ref{eomna}),
\begin{equation}
    \label{eq lV}
    \frac{1}{l_V^2} \sim e^2 f^2 \left(-X\right)^{n-1}.
\end{equation}
Let us now assume that the vector core is much smaller than the scalar one, $l_V\ll l_H$.
From (\ref{eq lV}) using the estimates, $f\sim v\,l_V/l_H$ and $(-X)^n\sim\lambda v^4$ at
$r\sim l_V$, we find
\begin{equation}
    \label{lVless}
    l_V \sim \frac{1}{\left(e^2\e\right)^{1/4}}.
\end{equation}
Note, that the above result is valid if $l_V\ll l_H$, which
can be recasted as follows, using (\ref{lVless}):
\begin{equation}
    \label{validVH}
    \frac{e}{\sqrt\lambda}\;\e^{(n-1)/n}\gg 1.
\end{equation}
In the opposite case, $l_V\gg l_H$, the non-linearity in $X$
is negligible, i.e. one has to set $n=1$ in (\ref{eq lV}).
Then we immediately find the size of the vector core as
\begin{equation}
    \label{lVgreater}
    l_V \sim \frac{1}{ev}.
\end{equation}
Eq.~(\ref{lVgreater}) is valid for $l_V \gg l_H$, or
\begin{equation}
    \label{validHV}
    \frac{e}{\sqrt\lambda}\;\e^{(n-1)/2n}\ll 1.
\end{equation}
For us the most interesting case is $\e\gg 1$, as we will see later,
this corresponds to the regime when the non-linearity in $X$
become important. Taking into account our assumption (\ref{1mass})
we may summarize our results for $\e\gg 1$ as follows. The size of
the scalar core (with the restored physical units) is given by
\begin{equation}
  \label{lH1}
    l_H= \frac{\eta}{M^2}\left[\lambda\left(\eta/M\right)^4\right]^{-1/2n},
\end{equation}
and the size of the vector core is
\begin{equation}
    \label{lV1}
        l_V\sim \frac{1}{e\eta}.
\end{equation}
It is worth to note that the vector core in our model is
roughly as large as in the standard case. This is 
what one can naively expect from the action (\ref{act}):
The kinetic term for the vector field is unchanged as 
compared with the standard Lagrangian, so it is unlikely 
that the vector core varies much.
%
\subsection{Vortex' mass}\label{subs mass}
%
Another way to see how the parameter $\gamma$ (\ref{gamma}) appears
in the model is to make the change of variables in the action (\ref{S}) as follows:
\begin{equation}
  \label{changeV}
  \phi=v f,\quad x=L_H\,y, \quad A_\mu = \frac{B_\mu}{eL_H},
\end{equation}
with $L_H$ given by (\ref{LH}). Substituting the rescaling (\ref{changeV})
into (\ref{S}) we immediately find the functional of the energy density
(with the restored physical units):
\begin{equation}
    \label{Efunc}
    E= \eta^2 \left[\lambda\left(\eta/M\right)^4\right]^{(n-1)/n} \mathfrak{F}(n,\gamma),
\end{equation}
where
\begin{equation}
  \label{F}
    \mathfrak{F}(n,\gamma)=\int{\rm d}^2y \left[ \frac{\gamma}{4} F_{\mu\nu}F^{\mu\nu}
    + (D_i f)^{2n} +\left(\mid f \mid^2-1\right)^2 \right].\nonumber
\end{equation}
Is it important to note that the above expression is only
valid in the non-linear regime in $X$, i.e. when $X\gg1$ inside the
scalar core. In addition, we have to require that the vector core
is not larger than the scalar one, $l_V\ll l_H$; otherwise
the vector core is partly outside the scalar one and Eq.~(\ref{Efunc})
is inapplicable.

To find the
energy density of the vortex for particular parameters of the Lagrangian,
one needs to calculate the functional of the energy density (\ref{Efunc})
applied to a solution. 
An alternative way is to minimize this functional. All these
methods require numerical methods to involve. It is possible, however, to
roughly estimate the energy density of a vortex,
based on the results of the previous subsection \ref{subs structure}.

There are three different contributions to the energy density
of the vortex, each associated with different terms in the action
(\ref{S}): The kinetic energy of the scalar,
\begin{equation}
    \label{epsK}
    \e_s\equiv -K(X),
\end{equation}
the potential energy,
\begin{equation}
    \label{epsP}
    \e_{\rm pot}\equiv V(\phi),
\end{equation}
and the kinetic energy of the gauge field,
\begin{equation}
    \label{epsV}
    \e_V\equiv \frac 14 F_{ij}^2.
\end{equation}
First we note that
the energy density inside the scalar core associated with the kinetic term $K(X)$
is approximately equal to the potential energy:
\begin{equation}
    \label{eps1}
    \e_s\sim \epsilon_{\rm pot}\sim \e,
\end{equation}
while the energy density of the gauge field is given by
\begin{equation}
    \label{eps2}
    \e_V\sim F_{ij}^2\sim \frac{1}{e^2 l_V^4}.
\end{equation}
Using (\ref{eps1}), (\ref{eps2}) and taking into account
(\ref{lH}), (\ref{lVless}) and (\ref{lVgreater}) it is
easy to estimate the energy density of the vortex for different
forms of the kinetic term $K(X)$:
\begin{equation}
    \label{Eest}
    E\sim \left\{\begin{array}{lcl}
     \eta^2, & & n\leq 1, \\
     \eta^2 \left(\lambda v^4\right)^{1-1/n}, & & n>1.\\
    \end{array} \right. \,
\end{equation}
Notice that Eq.~(\ref{Eest}) is in agreement with Eq.~(\ref{Efunc}). 
For $n>1$ the non-linearity in $X$ is important for
both the scalar and the vector fields, thus (\ref{Efunc}) is directly applicable.
For $n<1$ the vector core spreads wider than the scalar one, so
in the region $r\gtrsim l_H$ the kinetic term takes the standard form,
therefore we set $n=1$ in (\ref{Efunc}) and arrive at Eq.~(\ref{Eest}).

An important consequence of Eq.~(\ref{Eest}) is that for a particular choice
of the non-canonical kinetic term (namely, $n>1$), the energy per unit length
of a vortex can be (much) larger than that for the standard vortex. 
The opposite is impossible: There is no Lagrangian that leads to 
vortices with small energy per unit length. One can understand this
as follows: Although the scalar core can be adjusted to have a small size,
(exactly as in the case of global defects \cite{kdefect}), the vector core
nevertheless spreads widely, with the configuration close to the
standard case. 
Thus the contribution of the vector field to the energy is roughly the same
as for an usual vortex, as Eq.~(\ref{Eest}) shows.
%
\section{Constraints on the parameters of the action}
\label{SConstraints}
%
Let us now discuss constraints on the parameters of the model.
In this section we will closely follow the similar consideration
for the case of global $k$-defects \cite{kdefect}
with necessary adjustments.

First of all we must satisfy the hyperbolicity condition.
Physically it means that small perturbations
on the background solution do not grow exponentially.
As applied to our problem, we have to check that the perturbed
Eqs.~(\ref{eom phi}), (\ref{eom A}) give hyperbolic
EoMs for the propagation of small perturbations.
Note that for small enough wavelengths the gauge
derivative $D_\mu$ is replaced by the partial derivative $\partial_\mu$.
The Eq.~(\ref{eom phi}) for high wave-numbers becomes the EoM for a global scalar
$k$-field. The hyperbolicity condition for perturbations for $k$-essence
field reads \cite{Halo,Rendall}
\begin{equation}
  \label{hyperbolicity}
  \frac{K_{,X}(X)}{2X K_{,XX}(X)+K_{,X}(X)}>0.
\end{equation}
It is easy to check that for the Born-Infeld-like kinetic term (\ref{BI})
the hyperbolicity condition (\ref{hyperbolicity}) is met for $X<1/2$, while
for the second example we consider, Eq.~(\ref{X3}),
inequality Eq.~(\ref{hyperbolicity}) is always true.

Meantime the EoM for the gauge field (\ref{eom A}) in the limit
of small wavelengths coincides with
the standard EoM for the normal electromagnetic field, since
the r.h.s of (\ref{eom A}) can be neglected in this limit.

Thus we have proved that the system of equations
(\ref{eom phi}), (\ref{eom A}) is hyperbolic provided that the
inequality (\ref{hyperbolicity}) holds, and therefore
there are no instabilities for small wavelengths.
It is worth to note that with the above argumentation
we have not proved the stability of the system for long wavelengths.
This problem, however, deserves a separate investigation and is not addressed in this paper.

\begin{figure}[t]
    \psfrag{x}[l]{$v$}
    \psfrag{y}[l]{$\ln \lambda v^4$}
    \psfrag{standard}[l]{standard}
    \psfrag{vortex}[l]{vortex}
    \psfrag{quantum}[l]{quantum}
    \psfrag{k-vortex}[l]{k-vortex}
    \includegraphics[width=0.4\textwidth]{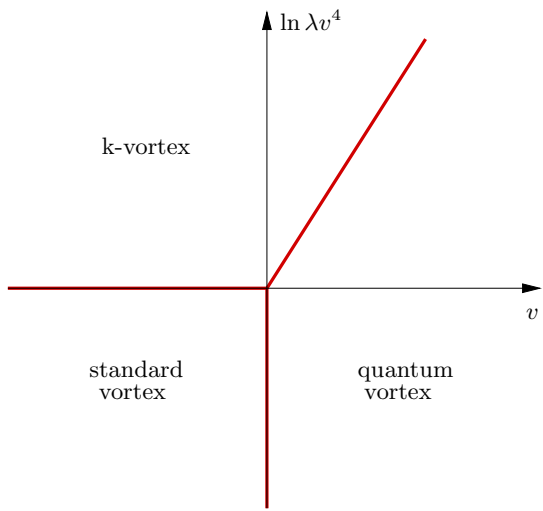}
    \caption{\label{constraint} Constraints on the parameters $\lambda$ and
    $v$ for the model (\ref{model}) are shown. There are three regions
    in the plane of parameters: i) standard vortex solution, when the
    non-linearity in $X$ inside the core is small and the standard solution
    is restored; ii) quantum defect, when the classical picture is not
    valid; iii) $k$-vortex, when the non-linearity in $X$ is large and
    the properties of a vortex are considerably different from the standard case.}
\end{figure}

As the second constraint on the parameters of the model
we demand that the nonlinear part
of $K(X)$ dominates inside the core of the defect.
Otherwise we end up with a ``standard'' solution arising
in the model with the canonical kinetic term. Thus we require
$  X\gtrsim 1$, which can be brought to
\begin{equation}
  \label{Xc}
  \lambda \left(\frac{\eta}{M}\right)^4\gtrsim 1.
\end{equation}
Finally, the third restriction comes from the validity of
the classical description. We consider vortices as classical objects,
neglecting quantum effects. This picture is valid if
the Compton wave length of the cube with the edge $l_H$
is smaller than the size of a scalar core $l_H$, and similar
must be true for the vector core.
This gives
\begin{equation}
    \label{quantum}
    l_H^4 \e_s \gtrsim 1,
\end{equation}
and
\begin{equation}
    \label{quantum2}
    l_V^4 \e_V \gtrsim 1.
\end{equation}
Eq.~(\ref{quantum}) can be rewritten as follows:
\begin{equation}
  \label{quant}
  \l \lesssim \e^{2-2/n},
\end{equation}
while (\ref{quantum2}) gives simply
\begin{equation}
  \label{quant2}
  e \lesssim 1.
\end{equation}
It is interesting to note that the only additional
constraint, as compared to the global $k$-vortices \cite{kdefect},
is a natural inequality Eq.~(\ref{quant2}).
We summarize the requirements (\ref{Xc}) and (\ref{quantum}) in
Fig.~\ref{constraint}.
%
\section{Numerical solutions}\label{Sec Numerics}
%
%
\begin{figure}[t]
    \psfrag{r}[l]{$r$}
    \psfrag{f}[l]{\large{$\frac{f(r)}{v},$}}
    \psfrag{a}[l]{$\alpha(r)$}
    \psfrag{canonical}[l]{canonical}
    \psfrag{DBI}[l]{DBI}
    \psfrag{cubic}[l]{power-law}
    \includegraphics[width=0.48\textwidth]{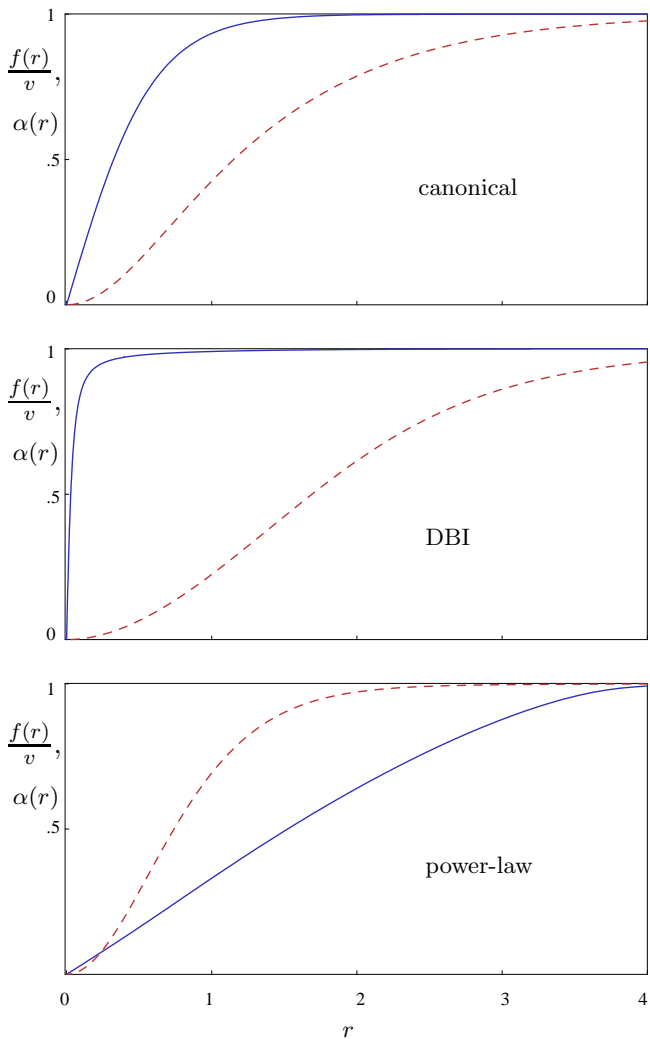}
    \caption{\label{Fig num} The numerical solutions for the field profiles
		$f(r)/v$ (solid), $\alpha(r)$ (dashed) are shown for different choice 
		of the kinetic term $K(X)$.		
		From the up to bottom: the standard case, $K(X)=X$; DBI term, $K(X)=1-\sqrt{1-2X}$;
		power-law term, $K(X)=X+X^3$. The parameters of the model are chosen such that the
		non-linearity in $X$ inside the core of a vortex is large, $\lambda=e=1/4$, $v=5$,
		The field profile for the vector part $\alpha$, 
		is roughly the same for the different kinetic terms, in accordance with (\ref{lV1}).
		While one can notice a strong dependence of the scalar field profile $f(r)$
		on the choice of $K(X)$. The size of scalar core is in a good agreement with our 			estimations (\ref{lH1}).}
\end{figure}
In this section we present the numerical solutions for the vortices
in the model (\ref{act}) [or, equivalently, (\ref{S})] with different choices of
the non-canonical kinetic term $K(X)$. We compare the obtained solutions
to the standard ones. With the help of these explicit solutions we verify our general
results on the properties of the gauge $k$-vortices, presented in Sec.~\ref{SGeneral}.

We solve numerically the system of ordinary differential
equation (\ref{EOM}), (\ref{EOMa}) for the model (\ref{S}) with
the following kinetic terms, $K(X)$:
\begin{itemize}
\item
canonical term, $K(X)=X$;
\item
DBI-like term, $K(X)=1-\sqrt{1-2X}$;
\item
the power-law kinetic term, $K(X)=X+X^3$.
\end{itemize}
In Fig.~\ref{Fig num} the functions $f(r)$ and $\alpha(r)$ are shown
for the vortex solution in the case of canonical, DBI and power-law kinetic terms.
We have chosen the parameters of the Lagrangian as $\lambda=e=1/4$ and $v=5$,
thus providing the non-linear in $X$
regime for the model with non-canonical terms (\ref{BI}) and (\ref{X3}),
since for these parameters $X\sim 10^2$.
One can see that the results of our general consideration, Sec.~\ref{SGeneral},
are in a perfect agreement with the numerical results
[compare Eqs.~(\ref{lH1}), (\ref{lV1}) with the numerical values
for the sizes of the scalar and vector cores].
The properties of $k$-vortices are indeed quite different
from those for a standard vortex in the regime when the
non-linearity in $K(X)$ is important. We also have found
the functions $f(r)$, $\alpha(r)$ for such parameters of the model,
that $X\leq 1$ inside the core of the defect.
As it was expected on general grounds, the obtained solutions
do not deviate much from the standard vortex solutions, since
in this regime
the kinetic terms (\ref{BI}), (\ref{X3}) have almost the canonical form.
%
\section{Summary and Discussion}\label{SDiscussion}
%
We have studied topological linear gauge defects
(gauge vortices), in the model with a non-canonical kinetic term.
The action for the model (\ref{act}) contains  kinetic terms
for the scalar and gauge vector fields and a symmetry-breaking potential.
The principal difference of the studied model from the standard one
is the presence of a non-standard kinetic part for the scalar field.
The term $K(X)$ in the action is in general some non-linear function
of canonical kinetic term $X$.

A remarkable feature of the model is that the
non-linearity of the kinetic term inevitably leads to the appearance
of a new scale in the action, a ``kinetic'' mass, in addition to
the ``usual'' scalar and vector masses.
The presence of another mass scale in the model changes radically the
basic properties of a vortex: the size of the scalar core and
the energy of a vortex per unit length vary considerably as compared to the standard case.

We have investigated general properties of $k$-vortices and
found restrictions on the parameters of the model, having in mind a rather
general form of a kinetic term with
the asymptotic behavior $K(X)\sim X^n$.
Also, for the sake of simplicity and clarity of results
we assumed that the scalar and vector
masses are of the same order, $e\eta \sim \sqrt\lambda\eta$.
We can summarize our general estimations as follows. The size of the scalar
core, $l_H$, depends on the coupling $\lambda$, and mass scales $\eta$ and $M$,
Eq.~(\ref{lH1}). A remarkable point is that $l_H$
roughly coincides with the characteristic size of the core in the
case of a global $k$-vortex \cite{kdefect}.
The size of the vector core $l_V$ does not depend 
on the kinetic mass $M$ and is roughly the same
as in the standard case, Eq.~(\ref{lV1}).

Having the values for the core' sizes one can
estimate the energy of $k$-vortex per unit length, see Eq.~(\ref{Eest}). 
An important result is that 
the mass of a vortex radically vary depending on the choice of kinetic term.
In the case $n>1$ and the limit $\lambda v^4\gg 1$, 
we have found a simple exact expression for the energy functional of $k$-vortex,
Eqs.~(\ref{Efunc}), (\ref{F}). 

As particular examples, we studied numerically two concrete models
having non-canonical kinetic terms: A DBI-like term, Eq.~(\ref{BI}), and
a power-law term, Eq.~(\ref{X3}).
The field profiles of domain walls for different choices of $K(X)$ are
shown in Fig.~\ref{Fig num}: The numerical solutions are in agreement
with our general estimations.

As we already discussed in our previous work \cite{kdefect},
interesting properties of $k$-defects
may have important consequences for cosmological applications.
Standard cosmic strings which might have been formed during phase
transitions in the early universe have a mass scale
directly connected to the temperature of a phase transition $T_c$,
$\mu\sim \eta^2 \sim T_c^2$.
By contrast, the mass scale of a resulting $k$-string depends
both on $T_c$ and the kinetic mass $M$.
This means that the tension of $k$-strings may not be close to $T_c^2$,
thus helping to avoid constrains on cosmic strings
$G\mu\lesssim 10^{-7}$ \cite{csc1,csc3} 
(or even stronger, $G\mu<3\times 10^{-8}$, see \cite{csc2}).
Meanwhile theoretical predictions give $G\mu\sim 10^{-6}-10^{-7}$ for GUT strings.
If, however, physics at the GUT scale involves
non-standard kinetic terms, then the GUT phase transition
may have lead to the formation of cosmic strings with smaller
tension, $G\mu \ll 10^{-6}$, thereby evading conflicts
with the present observations.
\begin{acknowledgments}
It is a pleasure to thank M.~Kachelriess for critical reading of the manuscript.
This work was supported by an INFN fellowship grant.
\end{acknowledgments}
%
%

\end{document}